\documentclass[prb,twocolumn,groupedaddress]{revtex4b4}
\usepackage{graphicx}
\begin{document}
\title{Entropic effects on the structure of Lennard-Jones clusters}
\author{Jonathan P.~K.~Doye}
\email{jon@clust.ch.cam.ac.uk}
\affiliation{University Chemical Laboratory, Lensfield Road, Cambridge CB2 1EW, United Kingdom}
\author{Florent Calvo}
\affiliation{Laboratoire de Physique Quantique,
IRSAMC, Universit\'e Paul Sabatier, 118 Route de Narbonne, F31062 Toulouse Cedex, France.} 
\date{\today}
\begin{abstract}
We examine in detail the causes of the structural transitions that occur for those small
Lennard-Jones clusters that have a non-icosahedral global minima. 
Based on the principles learned from these examples we develop a method to construct 
structural phase diagrams that show in a coarse-grained manner how the equilibrium 
structure of large clusters depends on both size and temperature. 
The method can be augmented to account for anharmonicity and quantum effects. 
Our results illustrate that the vibrational entropy can play a crucial role 
in determining the equilibrium structure of a cluster. 
\end{abstract}
\maketitle

\section{Introduction}

The determination of a cluster's equilibrium structure is a challenging task for the theoretician. 
For small clusters the first aim is usually to determine the lowest-energy
minimum on the potential energy surface using global optimization tools.
The global minimum must be the equilibrium structure at zero temperature, and it
is usually {\em hoped} that it will provide a good guide to the structure at higher temperatures. 
By locating the global minimum as a function of size a detailed picture of the 
cluster structure and the non-monotonic variation of cluster properties can be obtained. 
This can then allow a detailed comparison with experiments on size-selected clusters\cite{Parks97}
and mass spectroscopic studies where the abundances provide an indicator of 
particularly stable clusters.\cite{Harris84,Northby87,Martin96} 

However, global optimization becomes increasingly difficult as the size increases,
because of the exponential increase in the number of minima.\cite{Tsai93a,Still99}
Furthermore, optimization of some sizes may be particularly difficult because the potential
energy surface has an unfavourable topography.\cite{Doye99f,Leary00,Doye01c}
These effects limit the size for which global optimization is feasible.
For example, for the simple Lennard-Jones (LJ) potential, after extensive study and
computational effort, it is likely that the true global minimum for all LJ
clusters up to 150 atoms have now been found.\cite{Northby87,Deaven96,WalesD97,CCDLJ}
However, the last new putative global minimum in this size range was found 
as recently as 1999.\cite{Leary99}
Plausible putative global minimum have also been found up to $N$=309.\cite{Romero99,Hartke00}
For more complex potentials the present maximum feasible size is of course likely 
to be considerably smaller.

Some clusters already exhibit bulk-like structures in the sizes accessible by global
optimization (e.g.\ the alkali halides with rocksalt structure\cite{Twu90}), however many do not.
Extreme examples are provided by sodium clusters, which exhibit icosahedral structures
up to at least $20\,000$ atoms,\cite{Martin90} and by boron suboxide particles, 
which can have Mackay-like icosahedral structures with $10^{14}$ atoms.\cite{Hubert98} 

Therefore, to go beyond the maximum size feasible for global optimization a different
approach is necessary. This is possible if one settles for a more coarse-grained 
picture of the size evolution of cluster structure. 
So, rather than focussing on the the detailed size 
dependence, the general competition between structural types is instead followed.
The question is then what type of structure is most likely to be lowest in energy in a particular size regime,
rather than what is the particular arrangement of the atoms in the lowest-energy minimum.

Typically the energies of particularly stable sequences are compared to locate 
the crossover size at which the structure of the lowest-energy sequence changes.
This method is commonly used and so crossover sizes have been estimated for
LJ clusters,\cite{Xie,van89,Raoult89a}
metal clusters for elements such as tungsten,\cite{Marville87} nickel,\cite{Cleveland91} 
sodium,\cite{Wang} lead,\cite{Lim}, 
iron,\cite{Besley95} gold,\cite{Cleveland97b} calcium and strontium,\cite{Hearn97} 
rhodium and palladium,\cite{Barreteau00} and aluminium;\cite{Turner00} 
and molecular clusters for molecules such as  
N$_2$,\cite{Calvo99a} CO$_2$\cite{Maillet99} and SF$_6$.\cite{Boutin94}

To get accurate crossover sizes it is important that the lowest-energy sequences for each 
structural type are used. The shapes of the optimal sequences are those that most closely resemble the 
polyhedron derived from the Wulff construction (or a modified form 
of it for non-crystallographic structures\cite{Marks84}), which minimizes the surface free energy.
However too many of the studies using this approach compare sub-optimal 
clusters.\cite{Xie,Lim,Besley95,Hearn97,Barreteau00}
For example, it is particularly common to compare Ino decahedra and fcc cuboctahedra to Mackay icosahedra, 
because they can have exactly the same sizes, even though the most stable sequences are likely 
to more closely resemble truncated octahedra and Marks\cite{Marks84} decahedra.

As only the energies of the sequences are computed, the crossover sizes derived are
only strictly valid for zero temperature. Usually the possible effects of temperature 
on these crossovers are neglected, and it is {\em hoped} that they will provide a reasonable 
guide for higher temperatures. This neglect is probably because
a simple way to compute the entropic contribution to the free energy of 
a structural type has not been available.
Similarly in experiments on the evolution of cluster structure, 
size is usually the only variable that is considered,
and so a single (presumably temperature independent) crossover size is 
presented.\cite{Farges88,Kakar97,Torchet96,Reinhard97}

However, there is an increasing body of theoretical evidence that thermal effects can play 
a key role in determining a cluster's structure. There are a growing number 
of examples of solid-solid transitions in clusters where the structure changes from 
fcc or decahedral to icosahedral as the temperature increases.
Such transitions have been found to occur for those small LJ clusters that 
have a non-icosahedral global minimum,\cite{WalesD97,Doye98a,Doye98e,Doye99c,Leary99}
for gold clusters\cite{Cleveland98,Cleveland99}, for clusters interacting with a variable range 
Morse potential.\cite{Doye95c,BerryS00} and even for models of clusters of C$_{60}$ molecules.\cite{Calvo01d,Doye01e}
For molecular clusters, where low-pressure solid-solid or orientational phase transitions occur for bulk,
there has also been much interest in the size-dependence of the transition temperature. 
e.g. SF$_6$.\cite{Boutin94,Torchet95}

Here we examine in detail the effects of entropy on the equilibrium structure of LJ clusters. 
In Section \ref{sect:ss} we focus on the solid-solid transitions in those
LJ clusters that have a non-icosahedral global minimum. 
In particular we examine the different contributions to the entropy in detail.
Using the principles learned from these examples, in Section \ref{sect:phased}
we introduce methods to estimate the free energy of stable sequences of structures.
Thus we are able to construct structural phase diagrams showing how the 
structure depends upon both size and temperature.
There has been a previous attempt to calculate such structural phase diagrams based upon
macroscopic estimates of the different contributions to the free energy of a structure.\cite{Ajayan88}
However, it is unclear how valid the approximations made in that study are, in particular
the applicability of the macroscopic concepts to a small cluster. 
Clearly, a much more satisfactory approach is, as here, to compute 
the free energies directly from microscopic properties.
Finally in Section \ref{sect:discuss} we consider some of the 
other factors that determine cluster structure and how these effect the comparison
with experiment.
A brief report of some of this work has already appeared.\cite{Doye01b}

\begin{figure}
\begin{center}
\includegraphics[width=8.2cm]{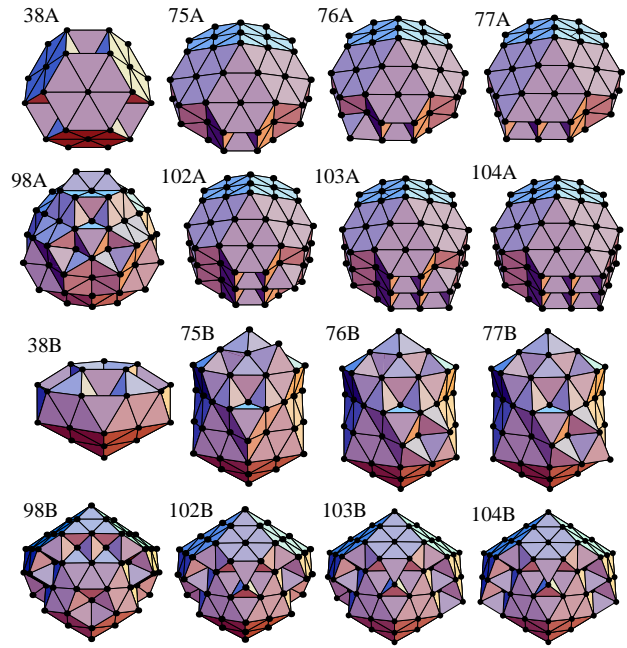}
\caption{\label{fig:nonicos} Structures of the global minimum (upper half) and 
the lowest-energy icosahedral minimum (lower half) for those LJ clusters with a 
non-icosahedral global minimum. The labels denote the size and energetic rank of the minima.}
\end{center}
\end{figure}

\section{Solid-Solid transitions}
\label{sect:ss}
The atoms in the clusters interact via a Lennard-Jones potential\cite{LJ} and
so the total potential energy is:
\begin{equation}
E = 4\epsilon \sum_{i<j}\left[ \left(\sigma\over r_{ij}\right)^{12} - \left
(\sigma\over r_{ij}\right)^{6}\right],
\end{equation}
where $\epsilon$ is the pair well depth and $2^{1/6}\sigma$ is the
equilibrium pair separation.

Most LJ global minima for $N<309$ have structures based upon the
series of Mackay\cite{Mackay} icosahedra.\cite{Northby87,Romero99}
However, there are a number of non-icosahedral global minima. 
These occur at sizes where the non-icosahedral morphology has a near optimum shape, 
whereas the icosahedral structures have an incomplete outer layer. 
For $N\le 147$ there are 8 such cases, which correspond to the 
fcc truncated octahedron at $N$=38,\cite{Gomez94,Pillardy,Doye95c} 
the Marks decahedra at $N$=75--77 and 102--104,\cite{Doye95c,Doye95d} and 
the Leary tetrahedron at $N$=98.\cite{Leary99}
These global minima and the lowest-energy icosahedral minima for these
sizes are depicted in Figure \ref{fig:nonicos}.
In the range $147<N\le 309$ there are a further eight non-icosahedral global minima
which are clustered in two sets around the complete Marks decahedra at $N$=192 and 238.\cite{Romero99,Hartke00}

For these examples it has been found that the global minima are not the most stable
structure up until the cluster melts, but that a low-temperature solid-solid transition
occurs at which icosahedral structures become thermodynamically more 
stable.\cite{WalesD97,Doye98a,Doye98e,Doye99c,Leary99}
However, these transitions are difficult to probe because there are large (free) energy
barriers between the structural types,\cite{Doye99c,Doye99f} making the dynamics of the transition extremely slow. 
They occur on time scales way beyond those accessible with conventional molecular dynamics.
For example, at the centre of the transition for LJ$_{38}$ the rate constant has been estimated to be
43s$^{-1}$ (using parameters appropriate for Ar).\cite{Miller99b} 
The equivalent rates for the larger clusters are likely to be orders of magnitude smaller.

The only simulation method that has been able to directly compute the 
equilibrium thermodynamics of the solid-solid transitions of any of these
clusters is parallel tempering \cite{Marinari92} (and its variants\cite{Calvo01a}).
Exchange of configurations between Monte Carlo runs at different temperatures
allows ergodicity to be achieved at the low temperatures relevant to the solid-solid transitions.
However, even for the smallest example, LJ$_{38}$, this method is computationally 
demanding.\cite{Neirotti00,Calvo00}

The superposition method \cite{Wales93a,Franke,Doye95a,WalesDMMW00} provides an alternative 
approach and one which is particularly useful for our present purposes because 
it allows us to analyse the different contributions to the entropy. 
In this approach the partition function is written as a sum over all the minima 
on the potential energy surface, i.e. $Z=\sum_i Z_i$, where $Z_i$ is the partition function for 
the basin of attraction surrounding minimum $i$. 
As long as all relevant minima are represented, ergodicity is ensured. 
The method is particularly simple to apply to solid-solid transitions as the number
of minima that contribute to the transition is comparatively few. 
The method can also be applied to the melting of clusters, 
but in this case a statistical representation of the distribution
of minima has to be used, because the number of minima associated with the liquid is, for all 
but the smallest clusters, too large to be enumerated.\cite{Doye95a} 
Another particularly useful feature of the method is that, by restricting the sum to a certain subset 
of the minima, the thermodynamic properties of a particular region of configuration space can be obtained,
i.e. $Z_A=\sum_{i\in A} Z_i$, where $Z_A$ is the partition function for region $A$. 

To apply this method to the solid-solid transitions in LJ clusters, we first 
generated large samples of low-energy minima for each cluster using established techniques for
searching potential energy surfaces.\cite{Tsai93a,Miller99a,WalesDMMW00} 
Some of these samples had been generated previously in work on the characterization of the 
potential energy surface  topography of LJ clusters using disconnectivity graphs.\cite{WalesMW98,Doye99f,Doye01c}
The samples included thousands of minima, which is far more than would actually 
significantly contribute to the low-temperature thermodynamics.
We then have to assume a form for the partition function of each minimum. 
Here, we use the harmonic approximation, which in the classical limit gives 
\begin{equation}
Z_i(T)={n_i \exp(-\beta E_i)\over (\beta h \overline{\nu}_i)^{\kappa}}
\label{eq:Zi}
\end{equation}
where $\beta$=$1/kT$, $\kappa$=$3N-6$ is the number of vibrational degrees of freedom,
$E_i$ is the potential energy of minimum $i$, 
$\overline{\nu}_i$ is the geometric mean vibrational frequency, and
$n_i$=$2N!/h_i$ is the number of permutational isomers of $i$,
where $h_i$ is the order of the point group. 

This harmonic superposition method (HSM) has been shown to give a correct qualitative 
picture of cluster thermodynamics (even of the melting transition \cite{Wales93a,Doye95a}), 
which becomes increasingly accurate at lower temperatures.
Accurate anharmonic forms for $Z_i$ are available,\cite{Doye95a,Calvo01b,Calvo01e} but the derivation
of the associated parameters characterizing the degree of anharmonicity adds an extra degree of complexity,
which is unnecessary here given the low temperatures at which the solid-solid transitions occur.

\begin{figure}
\begin{center}
\includegraphics[width=8.2cm]{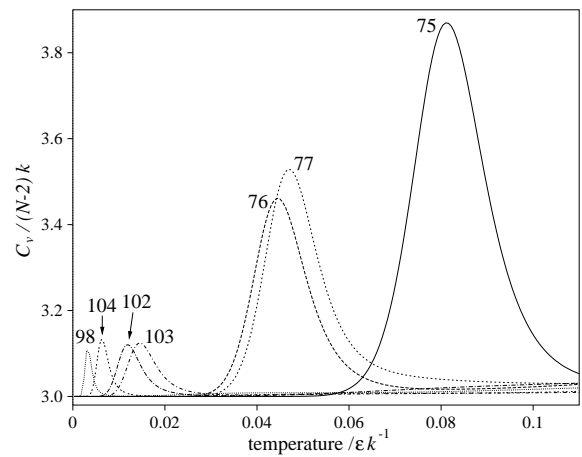}
\caption{\label{fig:Cv} Low temperature heat capacity peaks associated with the solid-solid 
transitions in those LJ clusters with non-icosahedral global minima. The curves are labelled by
the size of the cluster.}
\end{center}
\end{figure}

For LJ$_{38}$ the fcc to icosahedral transition has been previously shown to occur at roughly two-thirds of the
melting temperature and to give rise to a feature in the heat capacity.\cite{Doye98a,Doye98e,Doye99c,Neirotti00,Calvo00}
For the other clusters the transitions give rise to peaks in the heat capacity at low temperature, 
as shown in Figure \ref{fig:Cv}. For reference, melting transitions occur at roughly 
0.2--0.3$\epsilon k^{-1}$ for LJ clusters in this size range.\cite{Frantz95,Frantz01} 
Using the definition $Z_A=Z_B$ at the transition temperature, $T_{\rm ss}$, 
where $A$ and $B$ represent the two competing structural types, we have 
calculated $T_{\rm ss}$ values for all the clusters (Table \ref{table:Tss}).

\begin{table}
\caption{\label{table:Tss}Estimates of the solid-solid transition temperature $T_{\rm ss}$ 
and associated properties for those LJ clusters with less than 150 atoms that have 
non-icosahedral global minima.
$\overline{\nu}_A$ and $\overline{\nu}_B$ are the geometric mean vibrational frequencies 
and $h_A$ and $h_B$ are the orders of the point groups of the lowest-energy non-icosahedral 
and icosahedral minima, respectively. 
$n_A$ and $n_B$ have been evaluated using Equation (\ref{eq:nA}) at $T_{\rm ss}$
}
\begin{ruledtabular}
\begin{tabular}{cccccccc}
 & \multicolumn{3}{c}{$T_{\rm ss}/\epsilon k^{-1}$} & \\
\cline{2-4} 
 $N$ & HSM & Einstein & Eq.\ (\ref{eq:t_ss_est}) & $\Delta E/\epsilon$ & 
 $\overline{\nu}_A/\overline{\nu}_B$ & $n_B/n_A$ & $h_A/h_B$ \\
\hline
 38 & 0.121 & 0.199 & 0.316 & 0.676 & 1.0200 & 31.7 & 4.8 \\
 75 & 0.082 & 0.234 & 0.119 & 1.210 & 1.0475 & 91.8 & 20 \\
 76 & 0.046 & 0.223 & 0.053 & 0.510 & 1.0446 & 4.32 &  2 \\
 77 & 0.048 & 0.199 & 0.057 & 0.565 & 1.0451 & 6.56 &  4 \\
 98 & 0.004 & 0.009 & 0.006 & 0.022 & 1.0135 & 12.0 & 12 \\
102 & 0.013 & 0.096 & 0.014 & 0.086 & 1.0201 & 2.00 &  2 \\
103 & 0.016 & 0.116 & 0.018 & 0.107 & 1.0204 & 2.00 &  2 \\
104 & 0.007 & 0.069 & 0.008 & 0.048 & 1.0212 & 2.00 &  2 \\
\end{tabular}
\end{ruledtabular}
\end{table}

The transitions must clearly be driven by the greater entropy of the icosahedral structures,
but there are a number of contributions to this entropy. Firstly, there is the configurational
entropy due to the number of minima associated with a structural type. This has a component
due to symmetry. The number of permutational isomers decreases as the order of the point group 
of a minimum increases. This term particularly favours the icosahedral structures at $N$=38,
75 and 98, because of the high symmetry ($O_h$, $D_{5h}$ and $T_d$)
of the non-icosahedral global minima. 
The other contribution to the configurational entropy comes from the number of distinct low-energy
minima of a particular type. This term, of course, depends on temperature. At zero temperature only 
the lowest-energy minimum contributes to the partition function, $Z_A$, 
but as the temperature increases other low-energy minima begin to become populated. 
Furthermore, the contributions of other minima are greater the closer they are in energy 
to that of the lowest-energy minimum of that type. This term again favours 
the icosahedra because there are many low-energy icosahedral minima with different arrangements
of the atoms in the incomplete outer layer.\cite{Doye99f}

\begin{figure}
\begin{center}
\includegraphics[width=8.2cm]{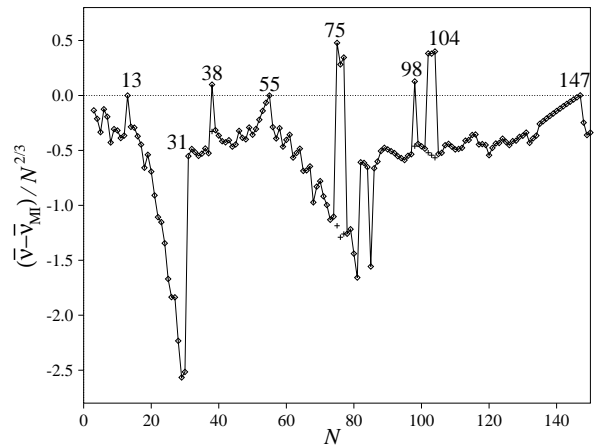}
\caption{\label{fig:vibgmin} Geometric mean vibrational frequency of the LJ global minima for
$N<150$. For those sizes with a non-icosahedral global minimum the value of $\overline{\nu}$ for
the lowest-energy icosahedral minimum is represented by a cross. The zero, $\overline{\nu}_{\rm MI}$, 
is a fit to the vibrational frequencies of the complete Mackay icosahedra using Equation (\ref{eq:nuvN}). }
\end{center}
\end{figure}

Secondly, there is the vibrational entropy. 
For the current examples the icosahedral structures have a smaller mean vibrational 
frequency (Table \ref{table:Tss}), which again favours the icosahedra. 
However, unlike the previous two contributions to the entropy, 
this term favours the icosahedra for a LJ cluster of any size, as illustrated by 
Figure \ref{fig:vibgmin}, where $\overline{\nu}$ (normalized to remove the non-specific
size-dependence) is plotted for the LJ$_N$ global minima.
Although there is a considerable variation in $\overline{\nu}$ for the icosahedral structures\cite{MaM}
(the complete Mackay icosahedra are the most rigid and those with an incomplete outer 
layer can have values considerably smaller than $\overline{\nu}_{\rm MI}$), 
only the vibrational frequencies for the non-icosahedral global minimum are greater than 
$\overline{\nu}_{\rm MI}$.

We can quantitatively analyse these contributions to the entropy and their effect on $T_{\rm ss}$.
We can neglect the differences in vibrational entropy by 
applying an Einstein-like approximation, where we assume that all the minima have the same 
mean vibrational frequency. The resulting values for $T_{\rm ss}$ are also 
given in Table \ref{table:Tss}. Although the transitions still occur, they do so
at significantly higher temperature with the error increasing with size. For 
the largest cluster the estimate of $T_{\rm ss}$ is roughly ten times too large.
This increasing importance of the vibrational entropy is because 
$\overline{\nu}$ is raised to the power $3N-6$ in the expression for 
the partition function of a minimum in Equation \ref{eq:Zi}.

We can quantify the effects of the configurational entropy, if we rewrite $Z_A$ as 
\begin{equation}
Z_A(T)={n_A \exp(-\beta E_A)\over (\beta h \overline{\nu}_A)^{\kappa}}
\label{eq:zA_single}
\end{equation}
where $n_A$, the effective number of minima (both permutational and geometric isomers)
associated with $A$, is given by
\begin{equation}
n_A(T)=\sum_{i\in A} {n_i \exp(-\beta (E_i-E_A))\over (\overline{\nu}_i/\overline{\nu}_A)^{\kappa}}.
\label{eq:nA}
\end{equation}
Using the properties of the lowest-energy minimum in region $A$ to define $E_A$ and $\overline{\nu}_A$, 
we give the relative values of the effective number of minima in each region at the solid-solid 
transition in Table \ref{table:Tss}. We also give the values of $h_A/h_B$, which correspond to the 
values of $n_B/n_A$ at zero temperature, where only differences in symmetry contribute.

As expected the effective number of icosahedral minima is greater. However, for the four largest clusters
$n_B/n_A=h_A/h_B$ at $T_{\rm ss}$ because,
although there are many low-energy icosahedral minima, the transition occurs at too low a 
temperature for them to be thermally populated, and thus to contribute to $n_B$.

\begin{figure*}
\includegraphics[width=17.6cm]{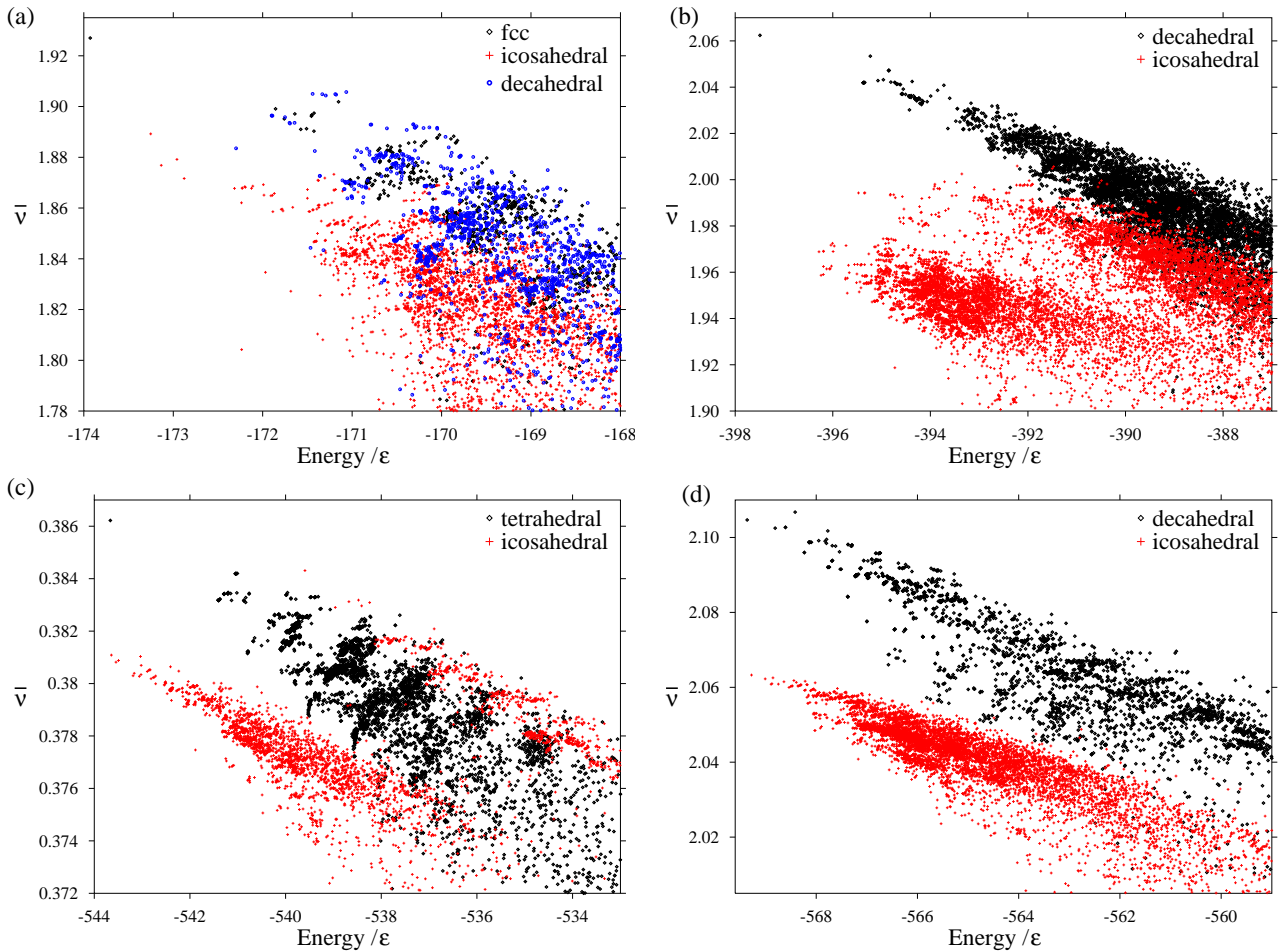}
\caption{\label{fig:vib_isomers}
Scatter plots showing how the geometric mean vibrational frequency of a minimum depends 
upon its energy and structure for (a) LJ$_{38}$, (b) LJ$_{75}$, (c) LJ$_{98}$ and (d) LJ$_{102}$.
Isomers with different structures have been differentiated using orientational order 
parameters.\cite{vanD92,Lynden95,Doye99c}
The unit of frequency is $(\epsilon/m\sigma^2)^{1/2}$.
}
\end{figure*}

From Equation \ref{eq:zA_single} we can obtain an expression for $T_{\rm ss}$: 
\begin{equation}
\label{eq:t_ss}
T_{\rm ss}={\Delta E\over k \left(\log\left(n_B/n_A\right)+
         \kappa\log\left(\overline{\nu}_A/\overline{\nu}_B\right)\right)},
\end{equation}
where $\Delta E$=$E_B-E_A$.
Of course, this equation cannot be straightforwardly applied to find $T_{\rm ss}$ because $n_B/n_A$ is 
a function of temperature. However, if we ignore the differences in configurational entropy,
this expression reduces to
\begin{equation}
\label{eq:t_ss_est}
T_{\rm ss}={\Delta E\over k \kappa\log\left(\overline{\nu}_A/\overline{\nu}_B\right)}.
\end{equation}
The values of $T_{\rm ss}$ that results from applying this expression are given in Table 
\ref{table:Tss}. The values become noticeably more accurate as the size increases.
This is because the vibrational term increasingly dominates over the configurational term in 
the denominator of Equation (\ref{eq:t_ss}). 

From this analysis we have shown that differences in vibrational entropy are crucial to the occurrence
of these solid-solid transitions. This conclusion is further strengthened by the absence of any 
solid-solid transitions when the LJ global minimum is icosahedral.
These results, therefore, suggest that solid-solid transitions, rather than being unusual,
should be expected for systems where different structural types have
systematic differences in $\overline{\nu}$ at sizes where the morphology with smaller vibrational
entropy is the lowest in energy.

\section{Structural phase diagrams}
\label{sect:phased}

Our aim is to produce a structural phase diagram showing how the LJ cluster 
structure depends on both size and temperature which is applicable to large clusters. 
At the relevant sizes (the energetic crossovers occur for clusters with thousands of atoms\cite{Raoult89a})
it is clearly not feasible to obtain large samples of minima as we have done for the examples considered
in the previous section. Methods are available to estimate the energy distributions of minima 
for large clusters,\cite{BerryS00} however these are involved and although possible for individual clusters
are not feasible for a wide range of sizes. 

Equation (\ref{eq:t_ss_est}) potentially provides a more readily
applicable approach, and as we have seen from the last section, this expression becomes more 
accurate as the size increases. 
Furthermore, its neglect of the differences in configurational entropy becomes more
realistic in the coarse-grained approach we are seeking. Although there are likely to be significant 
differences in the numbers of low-energy minima for morphologies $A$ and $B$ for many sizes (as for
some of the examples in the previous section), 
for some sizes $n_A>n_B$ but for other sizes $n_B>n_A$ and 
so on average they will be of the same order. 

Equation (\ref{eq:t_ss_est}) also ignores the potential variation of 
$\overline{\nu}$ for minima of the same structural type. 
Therefore, if we are to use Equation (\ref{eq:t_ss_est}),
this variation should be significantly smaller 
than the difference in $\overline{\nu}$ between structural types. 
Figure \ref{fig:vib_isomers} illustrates that this is true using a selection 
of the examples from the last section. 
The geometric mean frequency is a good order parameter to distinguish the structure of minima, and
the two bands of minima become increasingly distinct as the size increases. 

It is also noticeable that for both structural types in each diagram
there is a trend for the vibrational frequencies of the minima to linearly 
decrease as the energy increases. This is not unexpected. 
The higher energy minima tend to involve a less compact arrangement of the surface 
with fewer nearest-neighbour contacts and so have a lower frequency. 
The greater vibrational entropy of these minima is part of 
the reason why they are likely to become populated as the temperature increases. 
The creation of surface defects and increased surface diffusion result.\cite{Kunz93,Kunz94,Doye97b}
However, as the trends in $\overline{\nu}$ with energy are similar for minima of both 
structural types, the use of Equation (\ref{eq:t_ss_est}) is not compromised.

To use Equation (\ref{eq:t_ss_est}) to determine how crossover sizes depend on temperature
we need expressions for the size-dependence of $\Delta E$ and $\overline{\nu}_A/\overline{\nu}_B$.
Of course, it would be unfeasible to follow the detailed non-monotonic 
size variation of these properties over the size range we are interested in, because this would
require the global optimization of many very large clusters. Instead, we can use the properties of 
stable sequences of structures to construct a coarse-grained structural phase diagram.

The energies and vibrational frequencies of these stable sequences can be fitted to the forms
\begin{eqnarray}
\label{eq:EvN}
E(N)&=&a_E N +b_E N^{2/3}+ c_E N^{1/3} + d_E \\
\label{eq:nuvN}
\overline{\nu}(N)&=&a_{\nu} +{b_{\nu}\over N^{1/3}}+ {c_{\nu}\over N^{2/3}} + {d_{\nu}\over N}
\end{eqnarray}
where the first two terms represent volume and surface contributions, respectively.
These expressions can then be input into Eq.~(\ref{eq:t_ss_est}) to map out the solid-solid 
transitions in the structural phase diagram.
To complete the diagram, the melting line has to be determined.
Again it will have the form 
\begin{equation}
\label{eq:Tm}
T_m(N)=a_m +{b_m\over N^{1/3}}+ {c_m\over N^{2/3}} + {d_m\over N}.
\end{equation}

\begin{figure}
\includegraphics[width=8.2cm]{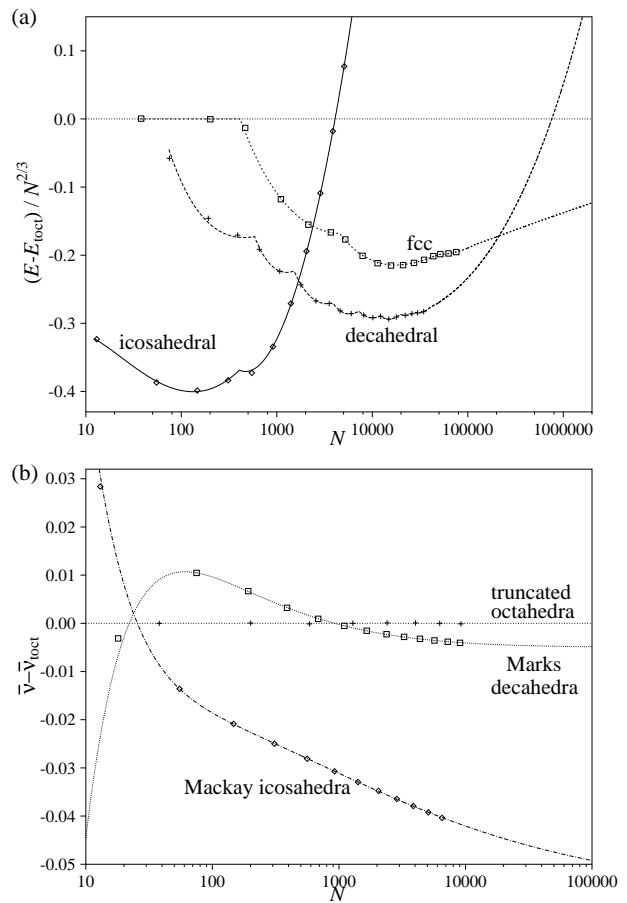}
\caption{\label{fig:Evib_large}
(a) Energies of the competing structural types for LJ clusters.
(b) Mean vibrational frequencies for Mackay icosahedra, Marks decahedra with grooves of depth one
and fcc truncated octahedra with regular hexagonal $\{111\}$ faces. 
In both parts, the data points represent specific clusters, the continuous lines
are fits to the data using Eq.~(\ref{eq:EvN}) and (\ref{eq:nuvN}), and
the zeroes, $E_{\rm toct}$ and $\overline{\nu}_{\rm toct}$, are a fit of the respective property for 
the fcc truncated octahedra with regular hexagonal $\{111\}$ faces.}
\end{figure}

The structural types that we should consider for LJ clusters are icosahedral, decahedral and fcc.
We have not considered unstrained close-packed clusters other than fcc. Generally, fcc clusters
will have a lower surface energy than hexagonal close-packed (hcp) clusters because it is easier to form
fcc structures with a higher proportion of $\{111\}$ faces. 
Therefore fcc clusters are usually favoured over hcp.
However, $a_E$ is slightly lower for hcp rather than fcc clusters due to a more favourable 
next-nearest neighbour shell, and so for sufficiently large sizes hcp clusters should be lower in energy.
Similarly, it can be energetically favourable to introduce twin planes into fcc clusters.\cite{Raoult89a}
The differences in energy, though, are very small and this behaviour is different from the rare gases which 
form fcc crystals. Therefore, we do not attempt to model this subtle structural competition.

\begin{figure}
\begin{center}
\includegraphics[width=8.2cm]{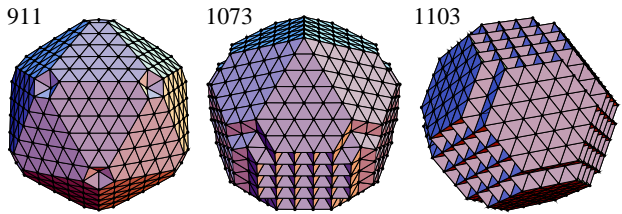}
\caption{\label{fig:large}
Examples of large icosahedral, decahedral and fcc clusters with optimal shape.
The sizes are as labelled.}
\end{center}
\end{figure}

The energies of the most stable sequences for icosahedral, decahedral and fcc clusters are 
represented in Figure \ref{fig:Evib_large}a, with the most stable sequences based on Mackay icosahedra, 
Marks decahedra and truncated octahedra, respectively. However, for each structural type there are small
changes in the optimal shape as the size increases, as forms that more closely
resemble the (modified) Wulff polyhedra become possible.\cite{Raoult89a} 
This gives rise to the lobes in each of the lines.
By $N$=549 it becomes favourable for the vertices of the Mackay icosahedra to be unoccupied.
As the size increases, $\{110\}$ facets of increasing size are introduced to make the 
Marks decahedral and truncated octahedral forms more rounded, 
and the grooves in the Marks decahedra become deeper.\cite{Raoult89a}
Three examples of structures that lie on the optimal lines are depicted in Figure \ref{fig:large}

Up to $N$$\approx$$10\,000$ the energies of these 
stable sequences of structures were obtained by minimization of all degrees of freedom.
Above this size minimizations were possible up to $N$$\approx$$35\,000$ for decahedral
and $N$$\approx$$80\,000$ for fcc structures if the cluster was constrained to maintain the 
correct point group symmetry ($D_{5h}$ for decahedral and $O_h$ for fcc). This constraint can reduce the 
computational cost by up to $h_i/2$ (i.e.\ 60 for $I_h$, 24 for $O_h$ and 10 for $D_{5h}$) 
while the errors in the energies of the minima arising from the use of symmetry are negligible.
These sizes are significantly larger than have been used in previous studies.\cite{Xie,van89,Raoult89a}

As the energetic crossover between decahedral and fcc structures lies beyond the largest size for which we 
could minimize the energy, we also had to use Equation (\ref{eq:EvN}) to obtain fits to the low-energy
envelope formed by the succession of optimal sequences, which would be valid to this crossover region.
The resultant energetic crossover sizes (Table \ref{table:xover}) 
are consistent with or an improvement upon the most comprehensive of the previous studies.\cite{Raoult89a}

In order to calculate $\overline{\nu}$ the Hessian matrix must be diagonalized. 
As the Hessian is not sparse---no cutoffs are applied to the potential so 
all atoms interact with each other---the cost of this operation scales as $N^3$.
This scaling limits the sizes for which $\overline{\nu}$ can be calculated.
It is unlikely that continuum elastic theory can provide a shortcut to obtaining 
the required vibrational information for large clusters, since, 
as a recent study of two-dimensional clusters showed, it only begins to give a good 
description of the lowest-frequency modes when the size is roughly 30 atomic diameters.\cite{Wittmer01}

We have been able to characterize the vibrational properties of the three sequences shown in 
Figure \ref{fig:Evib_large}b up to about $10\,000$ atoms, and other sequences up to about 3500 atoms.
As $a_{\nu}$ represents the volume contribution to the mean vibrational frequency, its value should be the same
for all sequences of a particular structural type. Therefore, we used our most well-characterized
sequences to determine $a_{\nu}$ for each structural type. 
This gave $a_\nu^{\rm icos}$=0.50970, $a_\nu^{\rm deca}$=0.51930, $a_\nu^{\rm fcc}$=0.52020.
For other sequences we used the appropriate $a_\nu$ value, 
and then fitted the three remaining parameters.

\begin{table}
\caption{\label{table:xover} The crossover sizes at zero temperature and
at the melting point for quantum xenon, argon and neon clusters, modelled using
an appropriate value of the de Boer parameter $\Lambda$, as compared
to classical Lennard-Jones clusters.
}
\begin{ruledtabular}
\begin{tabular}{cccccc}
 & & \multicolumn{2}{c}{$N_{\rm icos\rightarrow deca}$} & \multicolumn{2}{c}{$N_{\rm deca\rightarrow fcc}$} \\
\cline{3-4} \cline{5-6}
 & $\Lambda$ & $T=0$ & $T=T_m$ & $T=0$ & $T=T_m$ \\
\hline
 LJ & 0.000 & 1690 & 7440 &    $213\,000$ & $6\,670\,000$ \\
 Xe & 0.010 & 1810 & 6510 &    $269\,000$ & $5\,520\,000$ \\
 Ar & 0.030 & 2130 & 5500 &    $439\,000$ & $4\,380\,000$ \\
 Ne & 0.095 & 4640 & 6230 & $2\,844\,000$ & $6\,390\,000$ \\
\end{tabular}
\end{ruledtabular}
\end{table}

To construct the melting line $a_m$ is assigned the value of the zero pressure 
bulk melting temperature \cite{vanderHoef00} and the 
other three parameters in Eq.~(\ref{eq:Tm}) are fitted using the melting points
of the first four Mackay icosahedra, which were obtained from Monte Carlo simulations. 
For the 309-atom cluster parallel tempering proved to be necessary to ensure ergodicity. 
For this cluster a low-temperature shoulder was seen in the heat capacity, which was
indicative of a surface transition.\cite{Doye97b,Hendy01}

\begin{figure}
\begin{center}
\includegraphics[width=8.2cm]{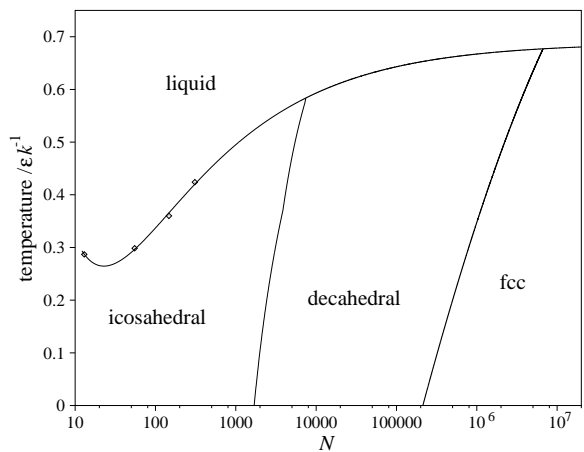}
\caption{\label{fig:phased}
Structural phase diagram for LJ clusters.
The data points represent the melting temperatures of the four smallest Mackay icosahedra. 
}
\end{center}
\end{figure}

The structural phase diagram that results from these calculations is shown in Fig.~\ref{fig:phased}.
The phase boundaries divide the plane into regions where the
majority of clusters have a particular equilibrium structure.
As the size increases, there is the expected progression from icosahedral to decahedral to fcc clusters.
The less-strained structures become favoured at larger $N$ as the effect of their lower
$a_E$ values dominates over their less favourable surface energies. 
For the sequences of Figure \ref{fig:Evib_large}b, 
$a_E^{\rm icos}$=$-8.5320$, $a_E^{\rm deca}$=$-8.6027$, $a_E^{\rm fcc}$=$-8.6101$.
Furthermore, the effect of the vibrational entropy can be clearly seen from the 
slopes of the phase boundaries in Fig.~\ref{fig:phased} and
from the differences between the crossover sizes at zero temperature 
and the melting point (Table \ref{table:xover}).
At higher temperatures icosahedra and Marks decahedra remain most stable up to considerably larger sizes 
than would be expected from their energetic crossovers. 
This is because of the relative values of their vibrational frequencies: 
$a_{\nu}^{\rm icos}$ and $a_{\nu}^{\rm deca}$ are 2.06\% and 0.19\% less than $a_{\nu}^{\rm fcc}$, respectively. 
Lower values of $a_{\nu}$ are associated with structural types with a greater strain energy; 
the physical origin of this correlation will be explored elsewhere.

Interestingly, the effect of entropy is greater for the decahedral to fcc transition,
even though the difference in $\overline{\nu}$ is smaller. 
The larger value of $N$ in the denominator of Eq.~(\ref{eq:t_ss_est}) and the much smaller difference in $a_E$ 
more than compensate for the smaller difference in $\overline{\nu}$.
For example, it can be easily shown that 
\begin{eqnarray} 
{dT_{\rm ss}\over dN}(T=0)&=&{\Delta a_E + 2\Delta b_E/3N^{1/3}+ 
                              \Delta c_E/3N^{2/3} \over k \kappa \log(\overline{\nu}_A/\overline{\nu}_B)} \nonumber \\
&\approx&{\Delta a_E \over k \kappa \log(\overline{\nu}_A/\overline{\nu}_B)} \qquad\hbox{for large }N,
\end{eqnarray}
where $\Delta a_E$=$a_E^B-a_E^A$, \dots. 

There has been much interest in the phase changes in clusters as the finite-size analogues 
of bulk phase transitions. The finite size, which can lead to 
unusual features, such as negative heat capacities,\cite{Labastie,Schmidt01a} causes any transition 
to occur not at a single temperature, but instead both phases can coexist over a range of temperature.\cite{Berry88} 
This is illustrated by the finite width of the peaks in Figure \ref{fig:Cv}.
We can easily estimate this coexistence range for our examples
if we define the upper and lower limit of this coexistence range, $T_{\rm ss}^\pm$, as the temperatures at
which $(p_A,p_B)$=(0.1,0.9) and (0.9,0.1). This gives
\begin{equation} 
T_{\rm ss}^\pm={\Delta E\over k \kappa\log\left(\overline{\nu}_A/\overline{\nu}_B\right)\mp \log 9}.
\end{equation}
However, as the size at which these transitions occur is large, the $\log 9$ term does not significantly
effect the denominator and so the coexistence range is very small.
When $T_{\rm ss}^\pm$ are plotted on Figure \ref{fig:phased} the lines are 
indistinguishable from those for the mid-point of the transition.

The assumption of harmonicity was one of the approximations used to construct 
the structural phase diagram of Figure \ref{fig:phased}. It has been previously
found that this approximation leads to an overestimate of roughly 10\% in
the value of the melting temperature for small LJ clusters, because the liquid-like 
state is significantly more anharmonic.\cite{Doye95a,Calvo01e}
One would expect the errors to be less for $T_{\rm ss}$ because the
differences in anharmonicity between two solid forms are likely to be much less 
and because the relevant temperatures are often much lower. 

One can introduce the effects of anharmonicity via temperature dependent 
frequencies, i.e.\ $\overline{\nu}(T)=\overline{\nu}^0(1-\beta^0/\beta)$,
where $\beta^0$ is a measure of the anharmonicity.\cite{Calvo01b,Calvo01e}
This then gives for the transition temperature:
\begin{equation}
{1\over k T_{\rm ss}^{\rm anharm}} = \frac{\kappa\left(\log(\bar \nu^0_A/\bar \nu^0_B) + 
                              \log\left({1-\beta^0_A/\beta\over 1-\beta^0_B/\beta} \right) \right) } 
                             {\Delta E}
\end{equation}
As the temperature occurs on both sides of this equation, 
it would have to be solved self-consistently. 
Alternatively an approximate solution can be obtained by taking
the first terms of the binomial expansion of the second log and then substituting in the harmonic
expression for $T_{\rm ss}$. This gives
\begin{equation}
{1\over k T_{ss}^{\rm anharm}} = \frac{\kappa\log(\bar \nu^0_A/\bar \nu^0_B)}{\Delta E}
+\frac{\beta^0_B-\beta^0_A}{\log(\bar \nu^0_A/\bar \nu^0_B)}
\end{equation}
The second term represents an anharmonic correction to $\beta_{\rm ss}^{\rm harm}$.
To apply this equation, estimates of the size dependence of $\beta^0$ would be required, 
which could then be fitted to a form equivalent to Equation (\ref{eq:nuvN}). 
Methods are available to obtain $\beta_0$,\cite{Calvo01e} but unfortunately we were 
not able to extend these to sufficiently large sizes to enable fits of sufficient accuracy to be 
generated that could be applied to the large sizes relevant to Figure \ref{fig:phased}.

We have also so far assumed that the clusters behave classically. 
However, it has recently been shown that quantum effects can have a significant effect 
on the thermodynamics of LJ clusters.\cite{Calvo01b} 
We can apply these same methods to obtain quantum structural phase diagrams, 
which are applicable to the rare gases. 
The quantum equivalent of Equation (\ref{eq:zA_single}) is 
\begin{equation}
Z_A(T)={n_A \exp(-\beta E_A^0)\over \prod_i^{\kappa}1-\exp(-\beta h \nu_A^i)},
\label{eq:zA_quantum}
\end{equation}
where $E_A^0=E_A+\kappa h\langle \nu_A\rangle /2$ includes the zero-point energy,
and $\langle \nu_A\rangle$ is the arithmetic mean frequency. 
To apply this equation to find $T_{\rm ss}$ it needs to be further simplified since 
it would require the characterization of the size evolution of the whole frequency distribution.
To illustrate the potential effects of quantum delocalization, we choose the 
simplest approximation, namely to represent
the distribution by $\kappa$ modes at the mean frequency.
Ignoring differences in the configurational entropy as before, this then gives
\begin{equation}
T_{\rm ss}^Q={\Delta E^0 \over k \kappa \log 
\left({1-\exp(-\beta h \overline{\nu}_A) \over 1-\exp(-\beta h \overline{\nu}_B)}\right)},
\end{equation}
where $\Delta E^0=E^0_B-E^0_A$.

\begin{figure}
\begin{center}
\includegraphics[width=8.2cm]{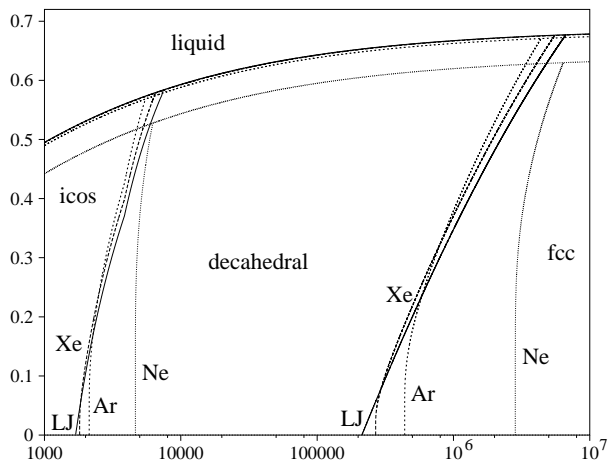}
\caption{\label{fig:quantum}
Structural phase diagrams for rare gas clusters, illustrating the effect of
the introduction of quantum effects. The classical results are solid lines (LJ), 
and the phase boundaries for the rare gas clusters are as labelled.
}
\end{center}
\end{figure}

The introduction of quantum effects can also lead to significant 
changes in the melting temperature of clusters.\cite{Neirotti00b,Calvo01b}
Following Ref.\ \onlinecite{Calvo01b} we simply model this change by
\begin{equation}
T_m^Q(N,\Lambda)=T_m(N)-\Lambda^2 \left(a_Q+{b_Q\over N^{1/3}} \right)
\end{equation}
where the coefficients $a_Q=5.18$ and $b_Q=6.84$ have been obtained using
results for small clusters,\cite{Calvo01b} and $\Lambda$ is the de Boer parameter.
$\Lambda$ is a measure of the quantum delocalization and corresponds to the value of 
$\hbar$ in reduced units, i.e.\ $\Lambda=\hbar/\sigma\sqrt{m \epsilon}$.

We have calculated structural phase diagrams for 
values of the de Boer parameter that are appropriate to some of the rare gases.
The results are illustrated in Figure \ref{fig:quantum} and Table \ref{table:xover}. 
There are two competing effects on the structural phase diagrams resulting from
the introduction of quantum delocalization.
Firstly, the zero-point energy stabilizes those structures with lower frequencies,
and therefore the zero temperature crossover sizes increase substantially as $\Lambda$ increases
(Table \ref{table:xover}).
Indeed the differences between neon and the other rare gases at low temperature are large enough 
to be potentially observable, if equilibrium clusters can be grown in this temperature range.
Secondly, for temperatures lower than the vibrational temperature $T_{\rm vib}=h\overline{\nu}/k$,
the vibrational contribution to the entropy is much less in the quantum case.
This is a corollary of the fact that for a quantum harmonic well $C_v/(N-2)$ is close to zero 
at low temperature and only rises near to 3 above $T_{\rm vib}$, whereas classically $C_v/(N-2)=3$ for all $T$.
Therefore, the temperature dependence of the crossover sizes are diminished, and the phase boundaries
initially rises vertically from the x-axis and in some cases even cross the classical line.

\section{Discussion}
\label{sect:discuss}

\begin{figure}
\begin{center}
\includegraphics[width=8.2cm]{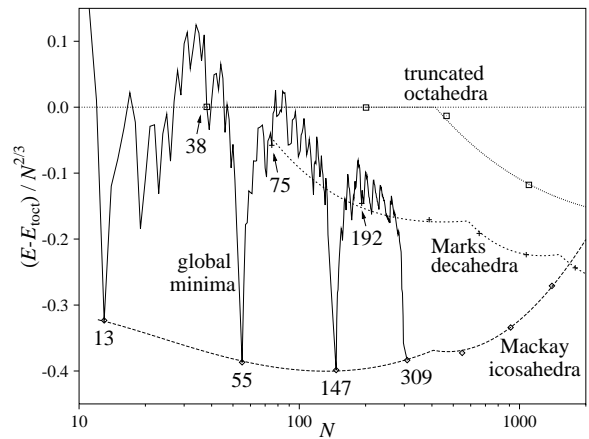}
\caption{\label{fig:Egmin} A comparison of the energy interpolated between particularly stable sequences 
and the energies of the global minima.
}
\end{center}
\end{figure}

The development of the structural phase diagrams presented here provides the starting
point for a more sophisticated discussion of the size evolution of cluster structure, 
which emphasizes the effect of both size and temperature. 
The one-dimensional picture which focusses on the effect of size and neglects thermal effects
should be discarded.
Therefore, we hope that this work will provide a stimulus for new experiments and calculations
that seek to illuminate the role of temperature in determining structure. 
Indeed, recent experiments on clusters of C$_{60}$ molecules\cite{Branz00} and 
strontium clusters\cite{Wang01} illustrate the interesting results that can 
be achieved. The methods developed here should be applicable to a wide range of materials, 
for which empirical potentials are available. An illustrative example for silver clusters 
modelled by the Sutton-Chen potential\cite{Sutton90} has already been provided.\cite{Doye01b}
This system has a structural phase diagram which was similar in form to that for LJ clusters.
However, it would be particularly interesting to map out the different possible families of 
phase diagrams. For example, in sodium clusters there would also be additional competition between
the icosahedral, decahedral and fcc structures considered here and clusters that are body-centred cubic.
These methods could also be potentially applied to clusters in a matrix or on a surface.

This point of view should also help in comparisons of experiment and theory, and of different 
experiments. Firstly, it helps one not to jump to hasty conclusions. For example, if two experiments
come to different structural conclusions for clusters of the same size, this does not 
necessarily imply that there has been a misinterpretation of the data; the differences
might simply reflect different temperatures.
Similarly, the failure of energetic calculations to reproduce the experimental observations
does not necessarily imply the inadequacy of the theoretical description of the interatomic forces,
because free energies need to be calculated for a rigorous comparison.
Secondly, and more positively, theoretical and experimental agreement on the size and temperature dependence
of structure would provide a strong vindication of the approaches used.

There are a number of aspects
that should be born in mind when interpreting the structural phase diagrams. 
Firstly, they only provide a coarse-grained picture of the size evolution, dividing
the phase diagram into regions where the {\it majority} of sizes have the same equilibrium structure.
To construct the diagrams we have assumed a monotonic dependence of properties on $N$ (Equations 
\ref{eq:EvN}, \ref{eq:nuvN} and \ref{eq:Tm}). However, as is well-known, most cluster properties
show a complex non-monotonic size dependence, particularly for smaller clusters.\cite{Jortner} Indeed, it 
is these size effects that are often of most interest.

This non-monotonicity is illustrated for $\overline{\nu}$ in Figure \ref{fig:vibgmin} 
and for $E$ in Figure \ref{fig:Egmin}.
Similarly, the variation in $T_m$ for small LJ clusters has been catalogued by 
Frantz.\cite{Frantz95,Frantz01}
It is noticeable that the variations in $E$ and $\overline{\nu}$ can be of a similar
magnitude to the systematic differences between the structural types.
For example, at three points the energies of the global minima in Figure \ref{fig:Egmin} coincide with those
for truncated octahedra ($N$=38) and complete Marks decahedra ($N$=75,192). Consequently, the 
real phase boundaries when resolved at the level of individual sizes will be much rougher than
in Figure \ref{fig:phased}. Indeed, the examples in Section \ref{sect:ss} correspond to small islands
of stability for fcc and decahedral structures, within a sea of icosahedral dominance at small $N$.

The relative importance of these variations does diminish with 
increasing size as the proportion of atoms in the surface decreases. 
Consequently, the coarse-grained approach used for the structural
phase diagrams becomes more reliable at larger sizes, but should be applied
with caution when crossovers occur for sizes with less than a few hundred atoms. 
There are other implications of this decrease in fluctuations.
For example, it is clear from Figure \ref{fig:Egmin} that the 38-atom 
truncated octahedron will be the only fcc global minimum up to at least $N$=1000.

Secondly, the position of the phase boundaries in the diagrams sensitively depends
on the, usually small, differences in the energetic and vibrational 
properties, in particular $\Delta a$ and $\Delta b$, between structural types.
It is asking a lot of an empirical potential to reproduce these properties
accurately enough to give phase diagrams that could agree in detail with experiments.
For the current example, where different structural types are all based in some sense upon close-packing 
but perhaps involving multiple twinning and hence strain, the crucial ingredients
for the energetics would be a correct reproduction of the relative surface energies
for $\{111\}$, $\{100\}$ and $\{110\}$ faces, and the elastic moduli that determine the
energetic penalty associated with the strain. 
Similarly, one would also want the potential to reproduce not only the bulk phonon dispersion curves, 
but also the effects of surfaces
and strains on the vibrational properties. Furthermore, if one were to apply this approach
to clusters, where, as for sodium, two different bulk structures could be adopted,
the potential would also have to be able to reproduce the properties of both forms.
However, most empirical potentials are only fitted to a small subset of these properties, 
usually those associated with the bulk.
Therefore, given the current limitations associated with empirical potentials and 
the current impossibility of performing ab initio calculations for the relevant size ranges, 
it is perhaps better to use the tools developed here to understand the qualitative form of the
structural phase diagram one would expect for a particular material.

Thirdly, the experiments one is seeking to compare with may not be at equilibrium.
Evidence for the non-equilibrium nature of many clusters has come from a number of sources.
It is suggested by the mixture of structures often observed for metal clusters, for example, 
when supported clusters are probed by electron microscopy\cite{Marks94} or x-ray diffraction;\cite{Zanchet00}
greater structural uniformity would be expected for samples of equilibrium clusters.
More rigorous evidence for the kinetic origin of this diversity comes from 
experiments that have shown that the structural composition can be changed simply 
through the cluster source conditions,\cite{Hall91,Reinhard97b}
and that well-defined structures are only obtained after careful annealing.\cite{Andres96}
Similarly, it has recently been shown for clusters of C$_{60}$ molecules
that the equilibrium non-icosahedral structures are only located after annealing at 
high temperature.\cite{Branz00,Baletto01c}

This difficulty of achieving equilibrium is also suggested by a range of recent theoretical results. 
For example, the potential energy surfaces of the examples in Section \ref{sect:ss} 
have a multiple-funnel character, where the different funnels correspond to 
the different structural types.\cite{Doye99f} 
This multiple-funnel character is not restricted to these examples 
but appears to be ubiquitous.\cite{WalesMW98,Doye99g,Doye01a,Doye01e} 
As the (free) energy barriers between funnels are typically large, the dynamics of 
structural transformations can be very slow, giving rise to the possibility of 
kinetic trapping in out-of-equilibrium structures. 

This effect is seen in recent growth simulations.\cite{Baletto00,Baletto01}
Typically, above a certain size, which is dependent on the material, temperature and growth rate,
structural transformations are unlikely, and so new growth then occurs around this 
preserved seed cluster.
The results are particularly dramatic for (C$_{60}$)$_N$ where perfect Mackay icosahedra are 
invariably formed, even though they are far from equilibrium.\cite{Baletto01c}
Similarly, it has been found that freezing of molten clusters give rise to a 
a mixture of products,\cite{Chusak01} where the relative amounts reflect
the width of the respective funnels,\cite{Doye00c} rather than the thermodynamics.
These results emphasise that it will be difficult to obtain equilibrium when a structural
type is only favoured at low temperature. This is particularly the case for the
larger sizes often associated with the decahedral to fcc crossover.

Many of the above considerations are relevant when we seek to compare our results 
for Lennard-Jones clusters with experimental results for rare gas clusters.
The first difficulty is that a clear-cut picture has not yet emerged from the experiments.
The most easy to interpret data comes from mass spectroscopic studies 
of rare gas cluster ions. Clear magic numbers that are signatures of icosahedral packing 
have been observed for small clusters of argon,\cite{Harris84,Harris86} krypton 
and xenon.\cite{Echt81,Miehle89} 
The icosahedra have been observed up to $N$=923, with no sign of any crossover 
to a different structural type.\cite{Miehle89} However, the situation is less clear for neon.
In the only mass spectroscopy experiment on neon clusters, of which we are aware, 
no strong signal of shell effects was found.\cite{Mark87}
It should also be remembered that there may be differences between the
structure of ions and neutrals, as illustrated by recent calculations for Ne$_N^+$,\cite{Naumkin98a} 
although these differences would be expected to be less important as the size is increased.

There have also been a series of experiments by Farges {\it et al.}\ using electron
diffraction to probe argon clusters.\cite{Farges83,Farges86,Farges88} 
To reach a structural assignment the experimental diffraction patterns have to be
compared to those calculated for structures of the appropriate size. 
Although this provides good evidence that small clusters are icosahedral with 
a suggested crossover from icosahedra at $N\approx 750$, it has
proved to be much harder to obtain a good fit to experiment for larger clusters 
with neither decahedral or fcc structures providing adequate agreement.
Van de Waal has investigated a series of models that provide better fits, 
which include features such as crossed stacking faults,\cite{van91,van93} 
crossed twins\cite{van96b,van96a} that could lead to faster growth.
For very large clusters, however, adequate agreement could only be obtained
from polycrystalline clusters, involving fcc, hcp and random close-packed domains.\cite{van00} 
These results again seem to confirm the difficulty of achieving equilibrium,
and are further reinforced by simulations of the freezing of LJ clusters, in which 
a mixture of structural forms is observed for larger clusters.\cite{Ikeshoji01}

Somewhat different conclusions have been reached from EXAFS experiments 
on argon clusters,\cite{Kakar97} in which fcc structures were suggested 
for clusters with more than 200 atoms. 
However, this conclusion was based on comparison with results
obtained for a very limited set of candidate structures, and 
no estimate of the experimental temperature was given. 

Given the above experimental situation it is not surprising that there is not yet
full agreement between experiment and theory. 
Furthermore, although the LJ potential provides a reasonable description
of the properties of the rare gases (including the phonon dispersion curves\cite{Dove}),
it has well-known deficiencies.\cite{Maitland} Therefore, one should not necessarily expect too close
agreement for the crossover sizes. However, we believe the basic form of the structural
phase diagram to be robust. Therefore, it would first be interesting to see
if an icosahedral to decahedral crossover could be revealed for argon clusters as the
experimental clusters are brought closer to equilibrium. Then, if this could be achieved, it
would be interesting to see if the temperature dependence of the crossover size that we 
have predicted here was reproduced.

The results we have obtained for LJ clusters have highlighted the important
role that differences in vibrational entropy can play in determining structure.
Striking examples for the considerable effect of vibrational properties on phase
diagrams have also been obtained for the thermodynamics of substitutional 
alloys.\cite{Ozolins01,Asta01,vandeWalle01}
Similarly, in the field of supercooled liquids there have been recent suggestions,
based upon simulation \cite{Sastry01} and calculations for model potential energy surfaces,\cite{WalesD01}
that vibrational properties could play a key role in the complex interplay 
that gives rise to the distinction between fragile and strong liquids. 

\acknowledgements
JPKD is grateful to Emmanuel College, Cambridge and the Royal Society for financial support.

\end{document}